\documentclass[12pt]{article}
\usepackage{graphicx}
\usepackage{amsmath}

\begin{document}

\title{Constraints on top-quark FCNC from electroweak precision
measurements}

\author{F. Larios$^a$\footnote{Also at Dept. of Physics and
Astronomy, Michigan State University.},
R. Mart\'{\i}nez$^b$, and M.A. P\'erez$^c$\\ 
a. Departamento de F\'{\i}sica Aplicada, CINVESTAV-M\'erida, \\
A.P. 73, 97310 M\'erida, Yucat\'an, M\'exico \\
b. Departamento de F\'{\i}sica, Universidad Nacional,\\
Apartado a\'ereo 14490, Bogot\'a, Colombia\\
c. Departamento de F\'{\i}sica, CINVESTAV,\\
A.P. 14-740, 07000 M\'exico, D.F., M\'exico}

\maketitle

\begin{abstract}

We study the one-loop contributions of the effective flavor changing neutral
couplings (FCNC) $tcZ$ and $tcH$ on the electroweak precision observables 
$\Gamma_Z, R_c, R_b, R_\ell, A_c$ and A$^{FB}_c$.
Using the known experimental limits on these observables,
we may place 95\% CL bounds on these FCNC couplings 
which in turn translate into the following limits for the branching ratios 
$BR(t \rightarrow cZ) \leq 6.7\times 10^{-2}$ and $BR(t \rightarrow cH) 
\leq (0.09 - 2.9)\times 10^{-3}$ for $114 \leq m_H \leq 170 GeV$. 

\end{abstract}
PACS numbers 14.65.Ha;12.60.C;12.15.M;12.15.L

\vspace{1cm}

As soon as it was confirmed that the flavor changing neutral couplings
(FCNC) of the top quark are highly suppressed in the standard model
(SM)\cite{lorenzo}, it was realized that some of its FCNC decay modes
can be enhanced by several orders of magnitude in scenarios beyond the SM,
and some of them falling within the LHC's reach\cite{review,jyang,aguilar}.
On the other hand, the use of effective Lagrangians in parametrizing
physics beyond the SM has been exploited extensively in FCNC top
quark couplings and decays\cite{weinberg,zhang1,zhang2}.
This formalism generates a model-independent parametrization of
any new physics characterized by higher dimension operators. In particular,
the use of this method to place limits on new physics effects by studing
their one-loop contributions to precisely measured observables has been
proved to be effective in the study of anomalous couplings of vector gauge
bosons \cite{wudka,toscano,dawson}, the top quark \cite{zhang1,yuan}
and the tao neutrino \cite{tesis}.  Under this approach,
several FCNC transitions have been also significantly constrained: 
$t \to c\gamma$ \cite{toscano,han}, $t \to cg$ \cite{han,gilberto},
$\ell_i \to \ell_j\gamma$ \cite{tesis,rodolfo} 
and $H \to \ell_i \ell_j$ \cite{loretos}.  The effective Lagrangian
technique has been also used to get limits on the scale $\Lambda$ 
associated to the new physics from the oblique parameters 
S,T,U \cite{zhang2,gabriel}.

In the present paper we are interested in getting the constraints imposed by
the electroweak precision observables $\Gamma_Z, R_c, R_b, R_\ell, A_c$ on the 
FCNC transitions $t\to cZ$ and $t\to cH$.
In order to perform a $\chi^2$ fit at 95\% CL, we will compute the one-loop
contributions of the tcZ/H couplings to $Z\rightarrow c\bar{c}$
which in turn will induce corrections to these observables. 

\begin{figure}[tbh]
\begin{center} 
\includegraphics[height=10cm]{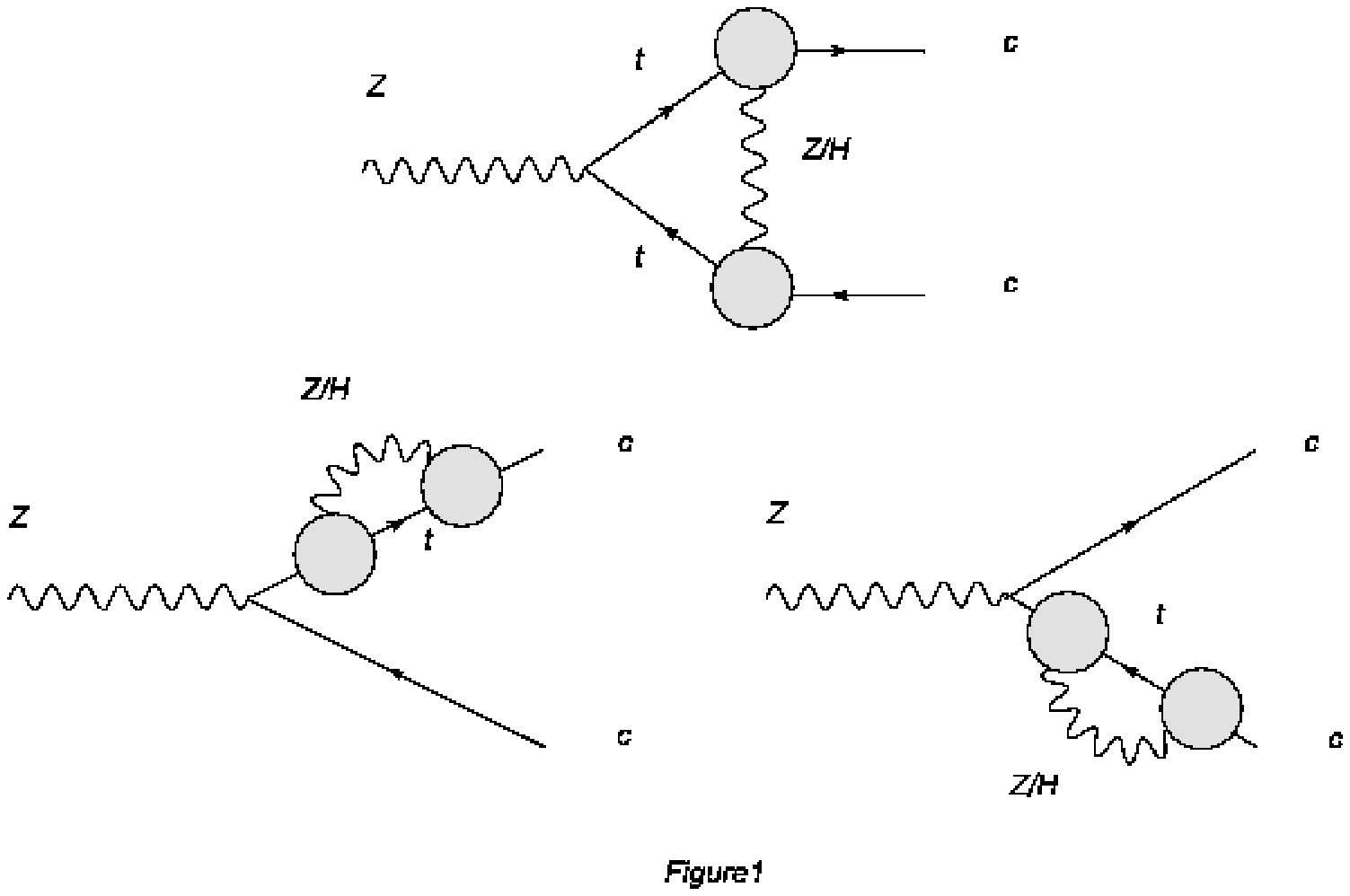} 
\end{center}
\caption{ Feynman diagrams for the one-loop contribution of the FCNC tcZ/H 
vertices to the decay mode $Z\rightarrow c\bar{c}$}
\label{figfeyn}
\end{figure}

We will use the following effective Lagrangian to parametrize
the FCNC of the top quark\cite{hewett}:
\begin{eqnarray}
{\cal L} = & &\bar t \{ \frac{ie}{2m_t} (\kappa_{tq'\gamma} + i\tilde 
\kappa_{tq'\gamma} \gamma_5)\sigma_{\mu\nu}F^{\mu\nu} \nonumber \\
&+& \frac{ig_s}{2m_t} (\kappa_{tq'g} + i \tilde \kappa_{tq'g}\gamma_5) 
\sigma_{\mu\nu} \frac{\lambda^a}{2} G^{\mu\nu}_a\nonumber \\
&+& \frac{i}{2m_t} (\kappa_{tq'Z} + i\tilde \kappa_{tq'Z}\gamma_5) 
\sigma_{\mu\nu}Z^{\mu\nu}  \nonumber \\
&+& \frac{g}{2c_w} \gamma_\mu(v_{tq'Z} + a_{tq'Z}\gamma_5) Z^\mu
\nonumber \\
&+& \frac{g}{2 \sqrt 2} (h_{tq'H} + 
i {\tilde h}_{tq'H}\gamma_5)H \}q^\prime
\end{eqnarray}
The one-loop contributions of the FCNC $tcZ$ and $tcH$ to the decay mode $Z
\to c\bar c$ are shown in Fig.~\ref{figfeyn}.
Even though the anomalous vertices enter in
these Feynman diagrams as a second order perturbation, we will see that the
known limits on the precision observables impose stringent constraints on the
couplings $tcZ/tcH$. This is not the case for the 
magnetic-dipole type couplings since their 
respective contributions are suppressed by an aditional $1/m_t$ factor. 

The width for the decay model $Z\to c\bar c$ may be expressed in the
following form after including the one-loop corrections.
\begin{equation}\Gamma(Z\to c\bar c) = 
\Gamma(Z\to c\bar c)_{SM} (1+ \delta^{Z/H}_{NP}),
\end{equation}
where the Z and H one-loop corrections are given by
\begin{equation}\delta^{Z/H}_{NP} = 
2 \left [ 
\frac{g^{SM}_V \delta g^{Z/H}_V + g^{SM}_A \delta g^{Z/H}_A} 
{(g^{SM}_V)^2 + (g^{SM}_A)^2} \right ], 
\end{equation}
with $g^{SM}_{V/A}$ the SM couplings of the Z gauge boson to the c quark 
and
\begin{eqnarray}\delta g^Z_{V,A} = Re[g^2_l F_L \pm g^2_r F_R],
\nonumber \\
\delta g^H_{V,A} = Re[h_l h_r (H_L \pm H_R)]. 
\end{eqnarray}
In the above expressions, we have used the definitions
\begin{eqnarray}g_{r/l}&=&\frac{i}{2c_W} (v_{tcZ} \pm a_{tcZ}),
\nonumber \\
h_{r/l} &=& \frac{1}{2\sqrt 2} (h_{tcH} \mp i {\tilde h}_{tcH}), 
\end{eqnarray}
and the functions $F_{L/R}$ and $H_{L/R}$ are given in terms of 
Veltman-Passarino functions and the dimensionless variables
$x_t = m_t/m_Z$ and $x_H = m_H/m_Z$, 
\begin{eqnarray}
F_L &=& \frac{g^2}{12\pi^2} \{ (-3+4s^2_w)\left[-3B_o(0,m_t, m_Z) + 
2B_o(m^2_Z, m_t, m_t)\right.\nonumber \\
&+&B_1(0, m_t, m_Z)+B_1(m^2_Z, m_t, m_t)\nonumber \\ 
&+& \left. x_t(B_o(0,m_t, m_Z) - B_o(m^2_Z, m_t, m_t))-
\frac{1}{2}\right] \nonumber \\
&+& x_t(12-12 s^2_W - x_t(3-4s^2_W)) C_o(x_t) \},\nonumber \\
F_R&=&\frac{g^2}{12\pi^2} \{ s^2_W \left[1+2B_1(0, m_t, m_Z) - 
2B_1(m^2_Z, m_t, m_t)\right.\nonumber \\
&+&\left. 2x_t (B_o(m^2_Z, m_t, m_t) - B_o(0,m_t, m_Z))\right]\nonumber \\
&+&(2s^2_W(3+2x_t - x^2_t) - 3x_t) C_o(x_t)\},
\end{eqnarray}
\begin{eqnarray}
H_L&=& \frac{g^2}{96\pi^2c_W} \{ (3-4s^2_W) \left[B_0(0,m_t, m_Z) - 
B_0(m^2_Z, m_t, m_t)\right.\nonumber \\                         
&+& B_1(m^2_Z, m_t, m_t) - B_1(0, m_t, m_Z)\nonumber \\
&+& (x_t-x_H-1)(B_o(0,m_t, m_Z) - B_o(m^2_Z, m_t, m_t))\nonumber \\
&+&\left. (x_t - x_H)^2 C_o(x_H) - \frac{1}{2} \right]+ 
\frac{4}{3} x_t s^2_W C_o(x_H)\},\nonumber \\
H_R &=& \frac{g^2}{96\pi^2c_W} \{ -4s^2_W \left[B_o(0,m_t, m_Z) - 
B_o(m^2_Z, m_t, m_t)\right.\nonumber \\
&-& B_1(m^2_Z, m_t, m_t) - B_1(0, m_t, m_Z)\nonumber         \\
&+& (x_t-x_H-1)(B_o(0,m_t, m_Z) - B_o(m^2_Z, m_t, m_t))\nonumber \\
&+& \left.((x_t - x_H)^2-x_t) C_o(x_H)\right]\nonumber \\
&+& 3 x_t C_o (x_H) \}.
\end{eqnarray}

We will use the above results to obtain 95\% CL limits on each individual 
anomalous coupling $tcZ/tcH$.  As mentioned before, from their
contribution to the effective  $Zc\bar c$ vertex shown in
Fig.~\ref{figfeyn} we can compute the deviation from the SM value of 
the following electroweak observables: 
\begin{eqnarray}
\Gamma_Z&=&\Gamma_Z^{SM}[1 + BR^{SM}(Z\to c\bar c)
\delta^{Z/H}_{NP}],\nonumber \\
R_c&=&R_c^{SM}[1 + (1 - R_c^{SM})\delta^{Z/H}_{NP}],\nonumber \\
R_b&=&R_b^{SM}(1 - R_c^{SM} \delta^{Z/H}_{NP}),\nonumber \\
R_\ell&=&R_\ell^{SM}(1 + R_c^{SM} \delta^{Z/H}_{NP}),\nonumber \\
A_c&=&A_c^{SM} (1 + \frac{\delta g_V^{Z/H}}{g^{SM}_V} + 
\frac{\delta g_A^{Z/H}}{g^{SM}_A} - \delta_{NP}^{Z/H}), \nonumber \\
A^c_{FB}&=&A^{c,SM}_{FB} (1 + \frac{\delta g_V^{Z/H}}{g^{SM}_V} + 
\frac{\delta g_A^{Z/H}}{g^{SM}_A} - \delta_{NP}^{Z/H}).\label{obser} 
\end{eqnarray} 

\begin{figure}[tbh]
\begin{center} 
\includegraphics[height=10cm]{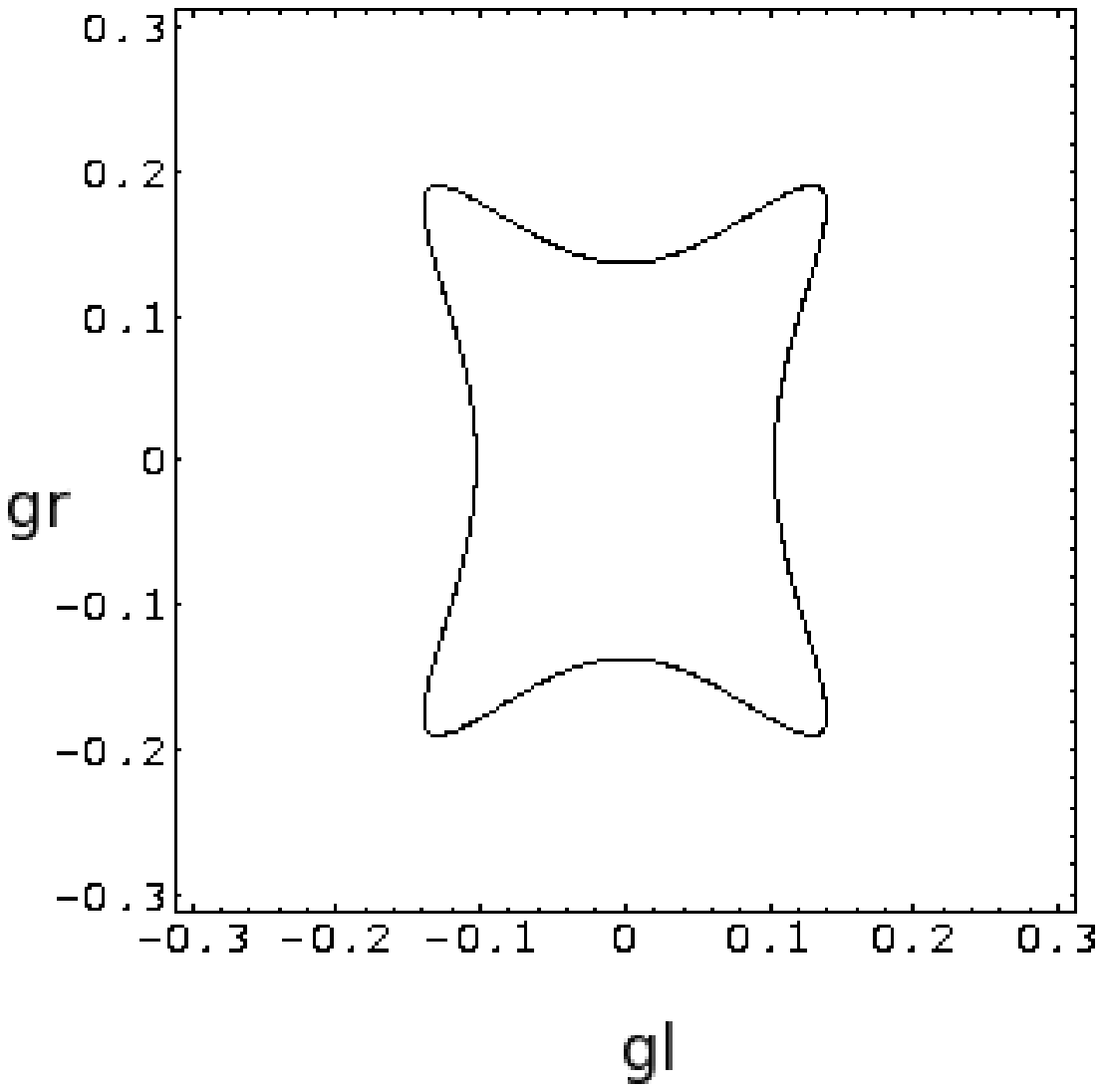} 
\end{center}
\caption{ A 95\% CL fit of the $tcZ$ coupling bounds obtained from the
current values for the electroweak precision observables
shown in Eq.(\ref{obser}). }
\label{figglgr}
\end{figure}

We now use the values given by the Particle Data Group \cite{pdg}
for the observables of Eq.~(\ref{obser}) and obtain the 95\% CL
limits for the $tcZ$ and the $tcH$ couplings.  We will neglect
possible interference effects and consider each coupling separately.
In Fig.~\ref{figglgr} the allowed parameter region in the
$g_l$-$g_r$ plane is shown.  We can compare with the recent limits
obtained by the DELPHI collaboration\cite{delphi}, from their
definition of the $tcZ$ coupling coefficient $\kappa_Z$.  We obtain
$\kappa_Z = 2 c_w \sqrt{g_r^2+g_l^2}$, and for $\kappa_\gamma=0$
DELPHI's upper limit $\kappa_Z \leq 0.4$ is equivalent to ours.
On the other hand, years ago a similar study on the $tcZ$ contribution
to FCNC processes like $B\to l^+l^- X$ put a stringent constraint
on $g_l$ ($\kappa_L \leq 0.05$)\cite{zhang1};  but not so much
for $g_r$ ($\kappa_R \leq 0.29$) for which the constraint came from
its contribution to the oblique parameters\cite{zhang1}.  
When computing the contribution of $tcZ$ to $b\to l^+l^- X$ one single
triangle diagram is considered (in the unitary gauge) where the
FCNC coupling appears only once, and this makes this process more
sensitive to the anomalous vertex.  
Our analysis puts similar bounds on the right handed $tcZ$ coupling
and it is based on a different set of variables than the ones
considered by Ref.~\cite{zhang1}.

\begin{figure}[tbh]
\begin{center}
\includegraphics[height=10cm]{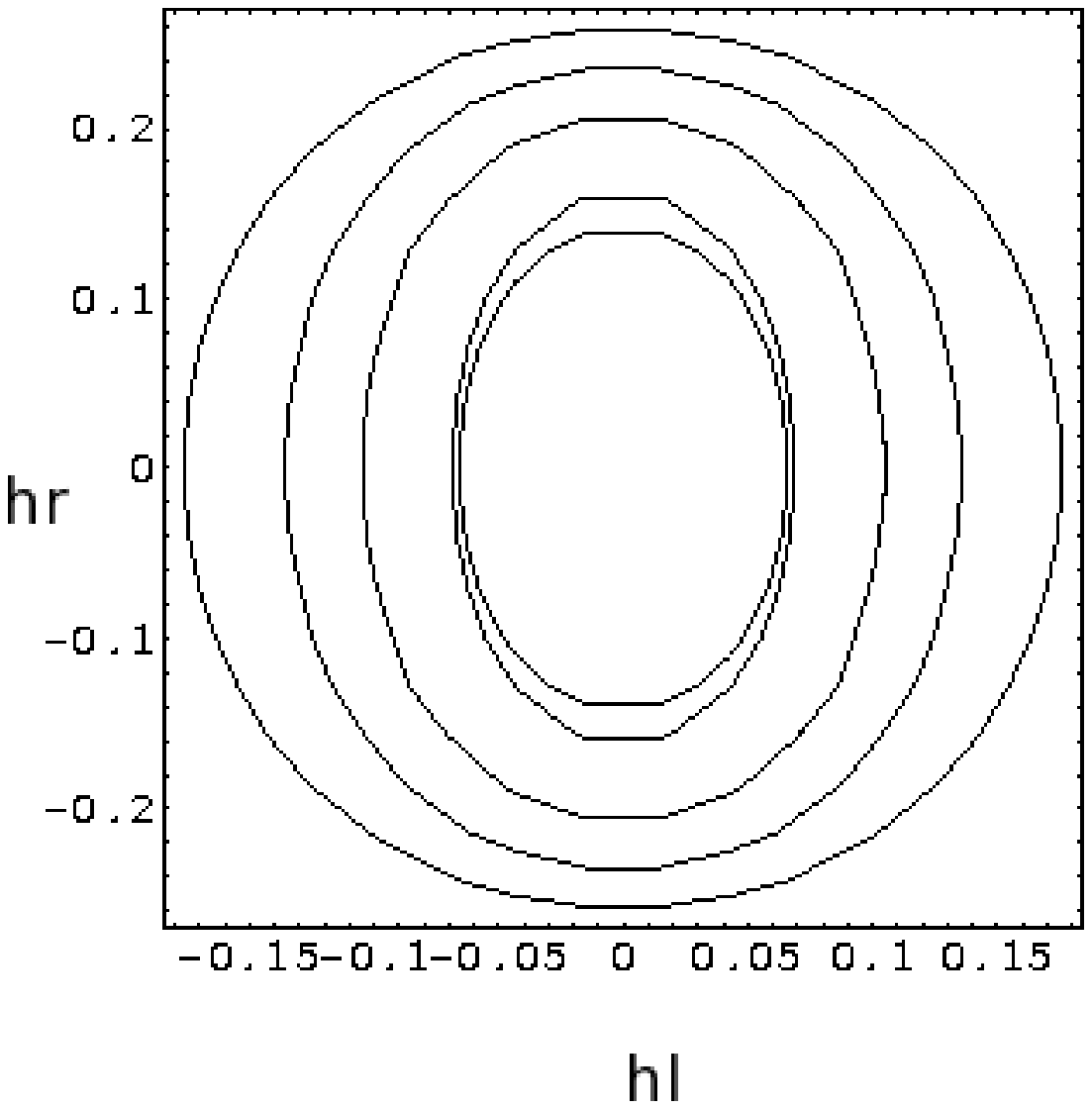} 
\end{center}
\caption{ A 95\% CL fit of the $tcH$ coupling bounds obtained from
the current values of the electroweak precision observables shown in
Eq.(\ref{obser}).  Upper limits from inner to outer contour line
correspond to the following values of the Higgs boson mass:
114, 130, 145, 160 and 170 GeV.}
\label{fighlhr}
\end{figure}

In Fig.~\ref{fighlhr} we depict the contours for
the 95\% CL upper limits on the $tcH$ coupling for
a selection of intermediate Higgs boson masses. 
The upper limits obtained for the 
$tcZ/tcH$ couplings can be translated into constraints on the respective 
branching ratios of the FCNC decay modes using the expressions 
\begin{eqnarray}\Gamma(t \to cZ)&=&
\frac{\alpha m_t(1-x_Z^2)^2(1+2x_Z^2) [g_l^2 + g_r^2]}
{8s_W^2x_Z^2}\nonumber \\
\Gamma(t \to cH)&=&
\frac{\alpha m_t}{8s_W^2} (1 - m_H^2/m_t^2)^2 [h_l^2 +h_r^2]
\end{eqnarray}
where $x_Z=m_Z/m_t$.

Finally, using the known expression for the SM decay width of the top
quark $\Gamma_t \cong \Gamma(t \to bW) = 1.6$GeV, and the limits of
Figs. \ref{figglgr} and \ref{fighlhr} we
obtain the following bounds on the FCNC decay modes of the top quark: 
\begin{eqnarray}BR(t \to cZ) &\leq & 6.7 \times 10^{-2},\nonumber \\
BR(t \to cH) &\leq & 0.9\times 10^{-4}, \;\; (m_H = 170 GeV)
\nonumber \\ 
BR(t \to cH) &\leq & 2.9 \times 10^{-3}, \;\; (m_H = 114 GeV).
\label{bratios}
\end{eqnarray}

As mentioned above, the upper limit on $BR(t\to cZ)$ is equivalent to
the recently reported by the DELPHI collaboration.
On the other hand, the limit on $tcH$ can be used as a test against
some possible beyond the SM contributions that could be of
order $10^{-3}$ to $10^{-1}$\cite{jyang}.
Some extensions of the SM with non-universal couplings to fermions can give 
sizeable $tcH$ couplings \cite{review,gilberto}.
In particular, it was found recently \cite{omar} that 
alternative Left-Right symmetric models with extra isosinglet heavy fermions 
may generate branching ratios for the $t \to cH$ mode as high as $2 \times 
10^{-3}$. Our bounds given in Eq.~(\ref{bratios}) clearly point to severe 
constraints on the parameters of this kind of models.

These bounds are also similar in size to the ones obtained within
the same approach for other FCNC top-quark decay modes:
$BR(t \to c\gamma) \leq 1.3 \times 10^{-3}$ and
$BR(t \to cg) \leq 3.4 \times 10^{-2}$.  Both obtained from the
observed $b \to s\gamma$ rate \cite{toscano,yuan,han}.

In conclusion, in this paper we have performed a systematic analysis of the 
radiate corrections induced on electroweak precision observables by the 
effective FCNC vertices $tcZ$ and $tcH$. We have found that at the 95\% CL 
the known values of these observables place significant constraints on the 
branching ratios allowed for the decay modes $t \to cZ, cH$ within the 
framework of the effective Lagrangian approach.

{\bf Acknowledgments}

We would like to thank CONACyT (M\'exico), Colciencias and
Fundaci\'on del Banco de la Rep\'ublica (Colombia) for support.

\newpage

\begin{thebibliography}{cc}
                                                            

\bibitem{lorenzo}
J.L. D\'{\i}az-Cruz, R. Mart\'{\i}nez, M.A. P\'erez, A. Rosado, 
{\it Phys. Rev. D}{\bf 41} (1990) 891; G. Eilam, J.L. Hewett, A. Soni, {\it 
Phys. Rev. D}{\bf 44} (1991) 1473; {\it erratum} Phys. Rev. D59 (1999)039901. 

\bibitem{review}
For reviews see, D. Chakaraborty, J. Konigsberg, D. Rainwater, 
{\it Annu. Rev. Part. Nucl. Sci.} {\bf 53} (2003) 301;
M. Beneke, I. Efthymiopoulos, M. Mangano, J. Womersley, et. al.,
{\it Top quark Physics: 1999 CERN Workshop on the SM Physics (and more)
at the LHC}, hep-ph/0003033.


\bibitem{jyang}
J.M. Yang, {\it Probing new physics from the top quark FCNC processes
at the linear colliders: a mini review}  Talk given at APPI2004,
Iweta, Japan, arXiv:hep-ph/0409351;
J. Cao, G. Liu and J.M. Yang arXiv:hep-ph/0311166.

\bibitem{aguilar}
J.A. Aguilar-Saavedra, {\it Top FCNC interactions: theoretical
expectations and experimental detection} Talk given at the final
meeting of the European Network ``Physics at Colliders'',
Montpellier, September 2004, arXiv:hep-ph/0409342;
J.A. Aguilar-Saavedra and G.C. Branco,
{\it Phys. Lett. B}{\bf 495} (2000) 347. 


\bibitem{weinberg} 
S. Weinberg, {\it Physica A}{\bf 96} (1979) 327; H. Georgi, {\it 
Nucl. Phys. B} {\bf 361} (1991) 339. 

\bibitem{zhang1} 
T. Han, R.D. Peccei and X. Zhang,
 {\it Nucl. Phys. B} {\bf 454} (1995) 527.

\bibitem{zhang2} 
R.D. Peccei S. Peris and X. Zhang,
 {\it Nucl. Phys. B} {\bf 349} (1991) 305;
R.D. Peccei and X. Zhang,
 {\it Nucl. Phys. B} {\bf 337} (1990) 269.


\bibitem{wudka} C. Artz, M.B. Einhorn. J. Wudka, 
{\it Phys. Rev. D}{\bf 49} (1994) 1370. 

\bibitem{toscano}
R. Mart\'{\i}nez, M.A. P\'erez, J.J. Toscano,
{\it Phys. Lett. B}{\bf 340} (1994) 91. 
                                                                          
\bibitem{dawson} S. Alam, S. Dawson, R. Szalapski, 
{\it Phys. Rev. D}{\it 57} (1998) 1577;
J. Bagger, S. Dawson, G. Valencia, {\it Nucl. Phys. B}{\bf 399} (1993) 364. 

\bibitem{yuan} R. Mart\'{\i}nez, J.A. Rodr\'{\i}guez, 
{\it Phys. Rev. D}{\bf 55} 
(1997) 3212; ibid. {\it Phys. Rev. D}{\bf 60} (1999) 077504; {\it Phys. Rev. 
D}{\bf 65} (2002) 017301; F. Larios, M.A. P\'erez, C.P. Yuan, {\it Phys. Lett. 
B}{\bf 457} (1999) 334; U. Baur {\it et al.}, arXiv:hep-ph/0412021.

\bibitem{tesis} F. Larios, R. Mart\'{\i}nez, M.A. P\'erez, 
{\it Phys. Lett. B}{\bf 345} (1995) 259. 

\bibitem{han} T. Han, K. Whisnant, B.L. Young, X. Zhang, 
{\it Phys. Rev. D}{\bf 55} (1997) 7241. 

\bibitem{gilberto}
A. Cordero-Cid, M.A. P\'erez, G. Tavares-Velasco, J.J. Toscano, 
{\it Phys. Rev. D}{\bf 70} (2004) 074003.

\bibitem{rodolfo} R.A. D\'{\i}az, R. Mart\'{\i}nez, J.A. Rodr\'{\i}guez, 
{\it Phys. Rev. D}{\bf 64} (2001)033004. 

\bibitem{loretos} J.L. D\'{\i}az-Cruz, J.J. Toscano,
{\it Phys. Rev. D}{\bf 62} (2000) 116005. 

\bibitem{gabriel} G. S\'anchez-Col\'on, J. Wudka, {\it Phys. Lett. B}{\bf 432} 
(1998) 383; P. Bamert et al., {\it Phys. Rev. D}{\bf 54} (1996) 4275. 

\bibitem{hewett} T. Han, J.L. Hewett, {\it Phys. Rev. D}{\bf 60} (1999) 074015.


\bibitem{delphi} Abdallah, J. et al, {\it Phys.Lett. B}{\bf 590} (2004) 21.

\bibitem{pdg} Particle Data Group, {\it Phys.Lett. B}{\bf 592} (2004) 1;
G. Altarelli, arXiv:hep-ph/0405182. 

\bibitem{omar} R. Gait\'an, O.G.Miranda, L.G. Cabral-Rosetti, arXiv:hep-
ph/0410268.

\end{thebibliography}
\end{document}